\documentclass[runningheads]{llncs}
\usepackage{graphicx}
\usepackage{tikz}
\usepackage[hyphens]{url}
\usepackage{hyperref}
\hypersetup{colorlinks=true, allcolors=black, breaklinks=true}
\usepackage{enumitem}
\usepackage{array, caption, floatrow, tabularx, makecell, booktabs}%

\DeclareRobustCommand\oneway  {\tikz[baseline=-0.6ex]\draw[->, thick] (0,0)--(0.5,0);}
\DeclareRobustCommand\twoway  {\tikz[baseline=-0.6ex]\draw[<->, thick] (0,0)--(0.5,0);}
\DeclareRobustCommand\dotted{\tikz[baseline=-0.6ex]\draw[->, thick,dotted] (0,0)--(0.54,0);}

\DeclareRobustCommand\dashedtwoway{\tikz[baseline=-0.6ex]\draw[<->, thick,dashed] (0,0)--(0.54,0);}

\usepackage{tikz}
\usepackage[most]{tcolorbox}

\colorlet{colexam}{red!75!black}
\tcbset{
  base/.style={
    empty,
    frame engine=path,
    colframe=gray!5,
    sharp corners,
    title={RQ\thetcbcounter~Result},
    attach boxed title to top left={yshift*=-\tcboxedtitleheight},
    boxed title style={size=minimal, top=4pt, left=4pt},
    coltitle=colexam,fonttitle=\large\bfseries\sffamily,
  }
}
\newtcolorbox[use counter=rqcounter]{rqres}{%
  base,
  drop fuzzy shadow,
  coltitle=white,
  borderline west={3pt}{-3pt}{black!50},
  attach boxed title to top left={xshift=-3mm, yshift*=-\tcboxedtitleheight/2},
  boxed title style={right=3pt, bottom=3pt, overlay={
    \draw[draw=black!70, fill=black!70, line join=round]
      (frame.south west) -- (frame.north west) -- (frame.north east) --
      (frame.south east) -- ++(-2pt, 0) -- ++(-2pt, -4pt) --
      ++(-2pt, 4pt) -- cycle;
  }},
  overlay unbroken={
    \scoped \shade[left color=black!10!black, right color=black]
    ([yshift=-0.2pt]title.south west) -- ([xshift=-1.5pt, yshift=-0.2pt]title.south-|frame.west) -- ++(0, -6pt) -- cycle;
  },
}

\begin{document}
\title{\mbox{HITA}: An Architecture for System-level Testing of Healthcare IoT Applications}
\titlerunning{Healthcare IoT Test Infrastructure Architecture}
%

\author{Hassan Sartaj\inst{1}\orcidID{0000-0001-5212-9787} 
\and
Shaukat Ali\inst{1}\orcidID{0000-0002-9979-3519} 
\and
Tao Yue\inst{1}\orcidID{0000-0003-3262-5577} 
\and
Julie Marie Gjøby\inst{2}\orcidID{0009-0002-8657-5691} 
}
\authorrunning{H. Sartaj et al.}
%
\institute{Simula Research Laboratory, Oslo, Norway\\
\email{\{hassan, shaukat, tao\}@simula.no}\\
\and
Section of Welfare Technologies, Oslo Kommune Helseetaten, Oslo, Norway\\
\email{julie-marie.gjoby@hel.oslo.kommune.no} \\
}
\maketitle              
\begin{abstract}
System-level testing of healthcare Internet of Things (IoT) applications requires creating a test infrastructure with integrated medical devices and third-party applications. 
A significant challenge in creating such test infrastructure is that healthcare IoT applications evolve continuously with the addition of new medical devices from different vendors and new services offered by different third-party organizations following different architectures. 
Moreover, creating test infrastructure with a large number of different types of medical devices is time-consuming, financially expensive, and practically infeasible. 
Oslo City’s healthcare department faced these challenges while working with various healthcare IoT applications. To address these challenges, this paper presents a real-world test infrastructure software architecture (\mbox{\mbox{HITA}}) designed for healthcare IoT applications. 
We evaluated \mbox{\mbox{HITA}}'s digital twin (DT) generation component implemented using model-based and machine learning (ML) approaches in terms of DT fidelity, scalability, and time cost of generating DTs.
Results show that the fidelity of DTs created using model-based and ML approaches reach 94\% and 95\%, respectively. 
Results from operating 100 DTs concurrently show that the DT generation component is scalable and ML-based DTs have a higher time cost.


\end{abstract}
\section{Introduction}
Healthcare Internet of Things (IoT) applications follow a cloud-based architecture to create an interconnected network with various medical devices and third-party applications~\cite{fiedler2013iot}. 
The primary objective of developing healthcare IoT applications is to create a central access point for medical professionals, patients, hospitals, pharmacies, and caretakers to deliver efficient healthcare services. 
Failure to provide timely healthcare services may lead to financial and human life loss. 
Therefore, automated and rigorous system-level testing of healthcare IoT applications is essential to ensure their dependability.

This work is conducted with Oslo City's healthcare department~\cite{oslocity}, 
which is working with various industries to develop healthcare IoT applications to deliver patients with high-quality services. 
One of the primary objectives is to create a test infrastructure for the system-level testing of healthcare IoT applications. 
Such test infrastructure requires integrating physical medical devices (e.g., medicine dispensers) and third-party applications (e.g., pharmacies) with a healthcare IoT application.  
A major testing challenge is that healthcare IoT applications evolve continuously with the addition of new medical devices, new/updated medical services, and new third-party applications. 
Integrating several different types of medical devices from various vendors is time-consuming, costly, and not a practical solution. 
Moreover, each third-party application has a limit on the maximum number of allowed requests for a particular time interval. 
Testing healthcare IoT applications within the limitations of third-party applications is challenging.

Several architectures have been proposed in the literature for developing healthcare IoT applications~\cite{muccini2018self}. 
A few works also utilize architectures for various software testing activities, e.g., integration testing~\cite{muccini2004using}. 
Our work focuses on designing a test infrastructure architecture to facilitate automated system-level testing of healthcare IoT applications.

To address the above-mentioned challenges, this paper presents a real-world test infrastructure software architecture (\mbox{HITA}) designed for healthcare IoT applications. 
\mbox{HITA} includes the DT generation (DTGen) and test stubs generation (TSGen) components to handle the integration of medical devices and third-party applications. 
We evaluate \mbox{HITA}'s DTGen component implemented using model-based and machine learning (ML) approaches. 
For the evaluation, we use a medicine dispenser (named Medido~\cite{medido}) integrated with a healthcare IoT application as a part of the experimental apparatus provided by Oslo City. 
Our evaluation analyzes the fidelity of DT created using both approaches, the scalability of the DTGen component in operating 100 DTs, and the time cost involved in generating DTs. 
The evaluation results indicate that the fidelity of DTs created using model-based and ML approaches is 94\% and 95\%, respectively. 
Results from operating 100 generated DTs concurrently show that the DTGen component is scalable. 
Results also indicate that ML-based DTs have a higher time cost compared to model-based DTs. 
In the end, we describe our experiences and lessons learned from applying \mbox{HITA} in a real-world industrial context.

The remaining paper is organized as follows. The related work is discussed in Section~\ref{rws}. \mbox{HITA} is described in Section~\ref{arch}. Evaluation is presented in Section~\ref{evaluation}, and lessons learned are outlined in Section~\ref{learnings}. The paper's conclusion is given in Section~\ref{conclusion}.

\section{Related Work}\label{rws}
Many works are available related to architectures for healthcare IoT targeting various aspects. 
This includes; analysis of IoT architectures for healthcare applications~\cite{muccini2018self},  
design patterns for healthcare IoT~\cite{mezghani2017model}, 
architecture for intelligent IoT-based healthcare systems~\cite{catarinucci2015iot}, 
architectural design decisions for developing digital twins of IoT systems~\cite{malakuti2018architectural}, 
a tool for modeling IoT architectures~\cite{sharaf2018arduino}, 
health monitoring architecture for IoT systems~\cite{azimi2017hich}, 
an architecture for IoT-based remote patient monitoring~\cite{al2018internet}, 
a requirements-based healthcare IoT architecture~\cite{lindquist2019iotility}, 
distributed IoT architecture~\cite{alnefaie2019towards}, 
health data sharing architecture~\cite{nguyen2021bedgehealth}, 
analyze architecture for healthcare IoT system~\cite{pise2023enabling}, 
architecture for an IoT-based healthcare monitoring system~\cite{moosavi2015sea}, 
and architecture for blockchain-driven healthcare IoT systems~\cite{sharma2023ehdhe}. 
In contrast to the aforementioned works, our work focuses on developing a test infrastructure architecture for healthcare IoT applications.

Several works are also available targeting architecture-based testing. 
This includes; analysis of architecture's function in software testing~\cite{bertolino2013software}, 
architecture-based test criteria~\cite{jin2001deriving},
architecture-driven integration testing~\cite{muccini2004using}, 
an architecture for analyzing fault tolerance~\cite{brito2008development}, and 
reliability assessment using architecture models~\cite{brosch2011architecture}. 
Compared to the above-mentioned works, our work offers a distinct contribution by presenting a test infrastructure architecture for healthcare IoT applications, which is designed to facilitate automated system-level testing of such applications.

\begin{figure}
\includegraphics[width=\textwidth]{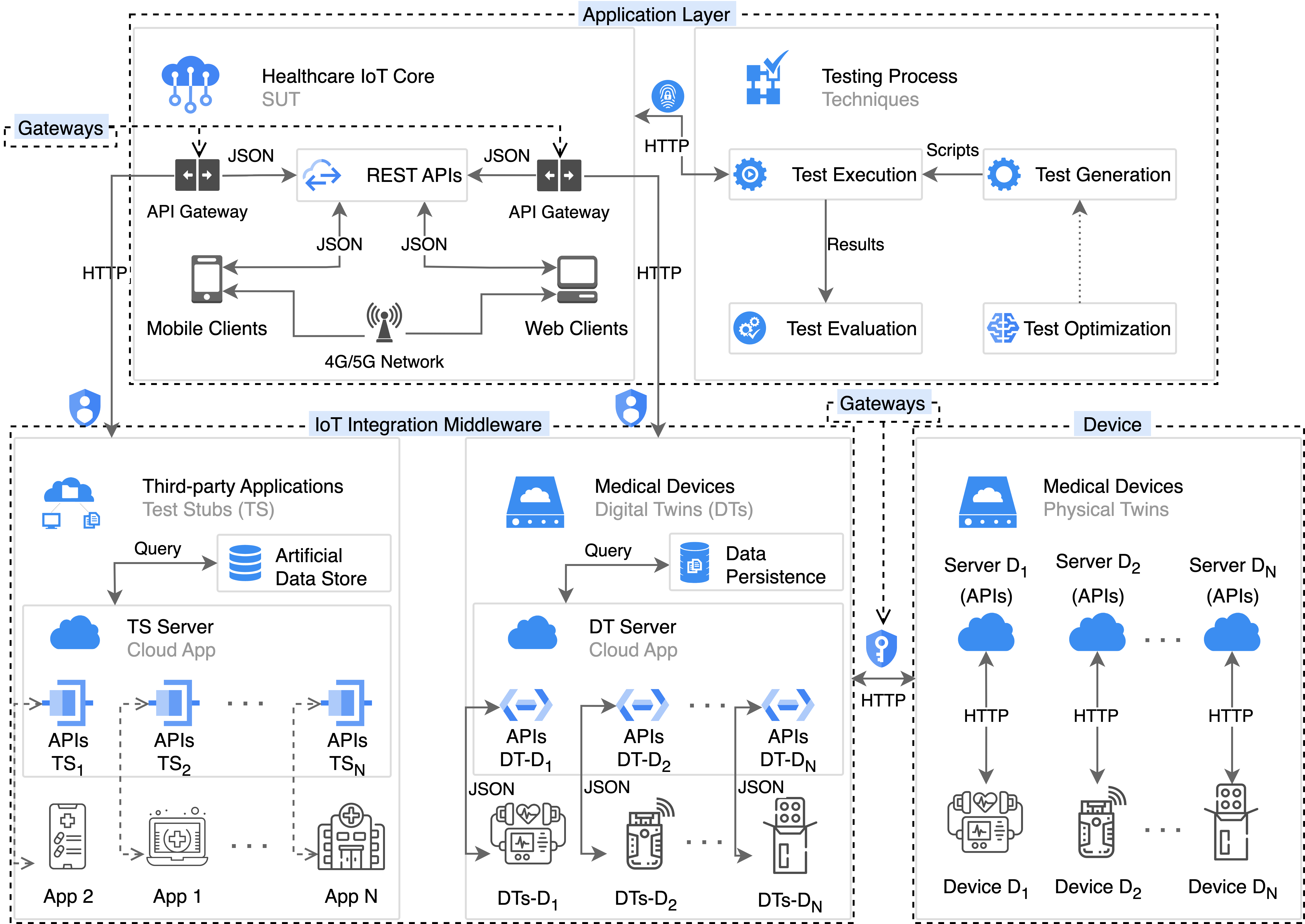}
\caption{An overview of \mbox{HITA}. The arrow (\oneway) shows one-way information flow, (\twoway) indicates two-way information flow, (\dashedtwoway) exhibits behavior simulation, and (\dotted) depicts optional information flow.} 
\label{fig:arch}
\end{figure}

\section{\mbox{HITA}: An Architecture for Test Infrastructure}\label{arch}

\subsection{\mbox{HITA} Components}
Figure~\ref{fig:arch} shows a real-world test infrastructure software architecture (\mbox{HITA}) designed for healthcare IoT applications. 
\mbox{HITA} is designed based on two commonly used architectural patterns, i.e., \textit{collaborative} and \textit{centralized} for healthcare IoT~\cite{muccini2018self}. 
\mbox{HITA} follows IoT reference architecture~\cite{guth2016comparison}, which is composed of an \textit{Application Layer} including healthcare IoT core and testing process, \textit{IoT Integration Middleware} with the DT generation (DTGen) and test stubs generation (TSGen) components, \textit{Gateways}, and \textit{Device} comprising physical medical devices.

\subsubsection{Healthcare IoT Core.}
The system under test (SUT) is a healthcare IoT application core that consists of several web and mobile clients for different users, such as patients, medical professionals, caregivers, and health authorities.
The primary communication channel for mobile clients (including iPads/Tablets) is the 4G/5G network due to its availability and access in remote areas. 
WiFi is used as an alternative communication channel in rare cases.
An important component of healthcare IoT applications is Application Programming Interfaces (APIs) developed according to the Representational State Transfer (REST) architecture~\cite{fielding2000architectural}. 
These REST APIs allow communication among various clients, third-party applications, and medical devices. 
The data interchange format used for this purpose is JavaScript Object Notation (JSON). 
To execute the tests on SUT, several different types of medical devices and third-party applications need to be integrated. 
\mbox{HITA} utilizes the DTGen component for medical devices and the TSGen component for third-party applications to handle integration challenges. 
API Gateways are used for secure communication with DTGen and TSGen components using Hypertext Transfer Protocol (HTTP) and secure API keys.

\subsubsection{Medical Devices - DTGen Component.}
For integration with medical devices, \mbox{HITA} utilizes the concept of digital twins to create a virtual representation of physical devices. 
Each medical device from a different vendor is connected to a server with several APIs for integration (as shown in the right-bottom of Figure~\ref{fig:arch}). 
The architecture for creating physical medical devices DTs consists of one \textit{DT Server} with APIs (e.g., \textit{APIs DT-D$_{1}$}) specific to a certain type of DTs representing a particular device (e.g., \textit{DTs-D$_{1}$}). 
APIs need to be developed following REST architecture~\cite{fielding2000architectural} to allow easy integration with SUT. 
DTs of medical devices can be generated with commonly used approaches~\cite{somers2022digital} like model-based approach or ML approach. 
In the case of multiple versions of a medical device, a separate DT for each device variant is required to be generated. 
For the communication between the \textit{DT Server} and DTs, JSON data interchange format is used. 
In case medical devices support different JSON schema, the Schema Registry can be used to ensure compatibility. 
The DTGen component also consists of \textit{Data Persistence} to preserve the state of DTs among various requests. 
The APIs of DTs are used to integrate DTs with SUT and physical medical devices. 
During testing, the DTs act as middle-ware between SUT and physical medical devices. 
DTs handle all communication traffic from SUT and communicate (via HTTP) with their physical twins when necessary.

\subsubsection{Third-party Applications - TSGen Component.}
To handle the challenges of integrating third-party applications for testing purposes, \mbox{HITA}'s TSGen component plays an important role. 
Each third-party application has dedicated servers with APIs for integration. 
The architecture for test stubs creation consists of one \textit{TS Server} with APIs  (e.g., \textit{APIs TS$_{1}$}) simulating the behavior of various applications (e.g., \textit{App 1}). 
The APIs for each test stub must be developed according to REST architecture~\cite{fielding2000architectural} for easy integration with SUT. 
Test stubs play a key role in replicating the functionality of third-party applications. 
For APIs requiring data (e.g., health data), the architecture includes an artificial data store with multiple databases corresponding to APIs representing different applications.  
The data manipulation is performed using query language compliant with the database type.

\subsubsection{Testing Process.}
The testing process starts with the test generation step using techniques for generating test data, test sequence, and test oracle.
Before testing SUT, it is important to ensure that DTGen and TSGen components adequately represent the desired behaviors. 
This can be done through pilot experiments evaluating the similarity in behaviors. 
The similarity in outputs should ideally be close to 100\% to have sufficient reliance on testing results. 
The generated tests in the form of test scripts are executed on the SUT.
Test execution requires API keys for communicating with SUT according to test scripts. 
The results of test execution are evaluated to analyze errors, faults, and failures. 
Moreover, test optimization during test generation is required for testing in a rapid-release environment and within a short time frame.

\subsubsection{\mbox{HITA} Operational Context.}
A tester initiates the testing process for testing a particular aspect of SUT, REST API testing, or graphical user interface (GUI) testing. 
This requires SUT to be integrated and operated with DTGen and TSGen components. 
Tests are executed on SUT through HTTP using JSON format. 
The SUT processes the request and communicates with medical devices DTs or third-party applications TS, depending on the test case. 
Finally, SUT generates a JSON response containing test execution results and sends it to the test execution module. 
This process continues for a specified testing budget.

\subsection{Quality Attributes}

\subsubsection{Scalability.} An important concern when testing a healthcare IoT application with a growing number of medical devices is \textit{scalability} of development efforts. 
\mbox{HITA} provides a component for digital twins used in place of physical medical devices during testing. 
Any number of digital twins corresponding to a particular medical device can be easily created and operated in \mbox{HITA}, either utilizing model-based or ML practices. 
Digital twins eliminate the physical need to integrate several medical devices and the risk of damaging physical devices. 
The use of digital twins is also \textit{cost-effective}, which is another key consideration for creating test infrastructure. 
In addition to digital twins, \mbox{HITA} has a \textit{Device} layer for connecting medical devices in the case physical devices are required in testing. 

\subsubsection{Maintainability.}
Using digital twins of medical devices and test stubs of third-party applications enables achieving \textit{maintainability} quality. 
Furthermore, \mbox{HITA} utilizes one server for DTGen and TSGen components that can operate locally or on the cloud, depending upon industrial preferences. 
Using one server each for both components requires less \textit{maintainability} effort as compared to using individual servers for different applications. 

\subsubsection{Extensibility.}
The modular structure of \mbox{HITA} components allows for achieving \textit{extensibility}. 
For each new medical device, digital twins and their APIs can be created using a model-based or ML approach. 
The APIs of digital twins are used for communication with SUT and the physical device. 
In the case of adding new healthcare services or features from a third-party application, a test stub is required to be created consisting of APIs for communication with SUT. 
The artificial dataset is created for testing if the new application is data-intensive and the data is unavailable or inaccessible. 

\subsubsection{Evolvability.}
\mbox{HITA} implicitly achieves \textit{evolvability} quality attribute by facilitating the evolvability of medical devices and third-party applications.  
Whenever a new medical device is added or an existing one is upgraded, the DTGen component can be utilized to generate a new DT or calibrate an existing one.
Similarly, when third-party applications undergo evolution, the TSGen component can be employed to generate or update test stubs.
This leads to achieving overall \textit{evolvability} requirements of SUT and integrated devices/applications during testing. 

\subsubsection{Heterogeneity.}
Creating test infrastructure for healthcare IoT applications involves integration with heterogeneous systems such as different medical devices and various types of third-party applications. 
\mbox{HITA} addresses this challenge by utilizing REST APIs, HTTP communication protocol, and the JSON data interchange format, providing a standardized method across all layers and components. 
This enables \mbox{HITA} to seamlessly support heterogeneous medical devices and third-party applications. 

\subsubsection{Security \& Privacy.}
Using real health records (e.g., patients' health data) during the testing process may lead to security breaches and data privacy issues. 
To handle security concerns, \mbox{HITA} imposes authentication and authorization mechanisms on all components. 
For data privacy, the TSGen and DTGen components consist of \textit{Artificial Data Store} and \textit{Data Persistence}, respectively, which contain synthetic data instead of real patients' health data. 

\subsubsection{Availability.}
A critical problem solved by \mbox{HITA} is the \textit{availability} of medical devices and third-party applications during testing. 
Running extensive tests with the goal of rigorous testing may lead to unavailable services. 
\mbox{HITA} uses third-party applications' test stubs in the form of APIs running on a server (\textit{TS Server}). 
Similarly, digital twins have \textit{DT Server} to handle requests during test execution. 
These servers are dedicated for testing purposes and hosted locally or on the cloud according to the up-time required for the testing process. 

\subsubsection{Robustness.}
The goal of testing a healthcare IoT application is to identify errors, faults, and failures with the assumption that integrated applications are robust. 
\mbox{HITA} instructs the development of APIs for DTGen and TSGen components following REST architecture, which provides a reliable mechanism for integration and communication among various applications~\cite{fielding2000architectural}. 
Moreover, as a result of test execution, DTGen, and TSGen components generate responses with failure and success information that enables identifying errors/faults in SUT.  

\subsubsection{Portability.} It is an additional feature of \mbox{HITA}. 
The architecture followed by TSGen and DTGen components can work on local machines when testing offline and remotely in different cloud environments.

\section{Evaluation}\label{evaluation}
We evaluate \mbox{HITA}'s DTGen component considering the DT fidelity, scalability of operating 100 DTs, and time cost of DT generation steps. 
We utilize model-based and ML approaches to generate DTs. 
We address the following research questions (RQs) in this evaluation.

\begin{itemize}[leftmargin=10pt]
    \item \textbf{RQ1:} \textit{What is the fidelity of DTs generated using model-based and ML approaches?}\\
    In this RQ, we analyze the fidelity DTs in terms of their similarity with corresponding physical devices. 
    \item \textbf{RQ2:} \textit{How do DTs operate with test infrastructure as the number of DTs scale up?}\\
    In this RQ, we assess the scalability of \mbox{HITA} by examining its ability to manage the operation of 100 DTs across various batch sizes (i.e., 10, 20, .., 100) during test execution. 
    \item \textbf{RQ3:} \textit{How much time is required in creating DTs using model-based and ML approaches?}\\
    In this RQ, we examine the time cost in various steps of model-based and ML approaches for creating DTs. 
\end{itemize}

\subsection{DTGen Component Implementation}
We generate DTs with two common approaches, i.e., model-based and ML approaches~\cite{somers2022digital}. The DTs generation steps of both approaches are described below.  

\subsubsection{Model-based DT.}
To generate model-based DT, we follow the approach presented in~\cite{sartaj2024modelbased}, which is briefly described below. 
In the first step, we model the structural and behavioral aspects of a medical device for which we need to create DTs.
This requires creating a domain model of a medical device to capture device concepts and properties, specifying constraints on device properties using Object Constraint Language (OCL), and modeling the device behavior in the form of state machines. 
Creating models and constraints in the first step is manual. The next steps are automated. 
In the second step, we create an instance model of the domain model using the input device property values (in JSON format) and validate the instance model using OCL constraints~\cite{sartaj2019search}. 
In the third step, we create a device state machine as the owned behavior of the device instance model. 
Finally, we make a device executable model to operate DT simulating the device.

\subsubsection{ML-based DT.}
Generating DTs using ML requires medical devices' data for training. 
Since medical devices are assigned to patients, their data is inaccessible. 
In the first step, we collect data using a REST API testing tool (i.e., EvoMaster~\cite{arcuri2019restful}). 
For each test execution, we collect data from the API request and response. 
In the second step, we preprocess the data to remove outliers and handle missing values. 
In the third step, we define a neural network (NN) architecture with an input layer representing total features and output layers representing total classes. 
The number of hidden layers, dropouts, and activation functions are finalized through a pilot experiment for hyperparameter search.  
In the fourth step, we train the NN with the training data. 
For training, we use the Adam optimizer and cross-entropy loss function, which are suitable for classification tasks in ML. 
At the end of training, we store the trained model. 
Finally, to operate a DT, we load the trained model and prepare for inference. 

\subsubsection{Development Utilities.}
We implemented the DTGen component in Python, with different frameworks and libraries employed for both DT generation approaches. 
For the model-based DTs, we utilized the PyEcore\footnote{https://github.com/pyecore/pyecore} framework to handle modeling aspects of DT. 
For ML-based DTs, we used Scikit-learn~\cite{scikit-learn} library and PyTorch framework. 
To develop a DT server for the DTGen component, we utilized the Flask\footnote{https://flask.palletsprojects.com/en/2.2.x/} framework, with Flask-RESTful library for creating REST APIs of DTs. 
In addition, to facilitate DTs' data persistence and communication, we used JSON data interchange format.

\begin{figure}
\centering
\includegraphics[width=10.9cm, height=4.4cm]{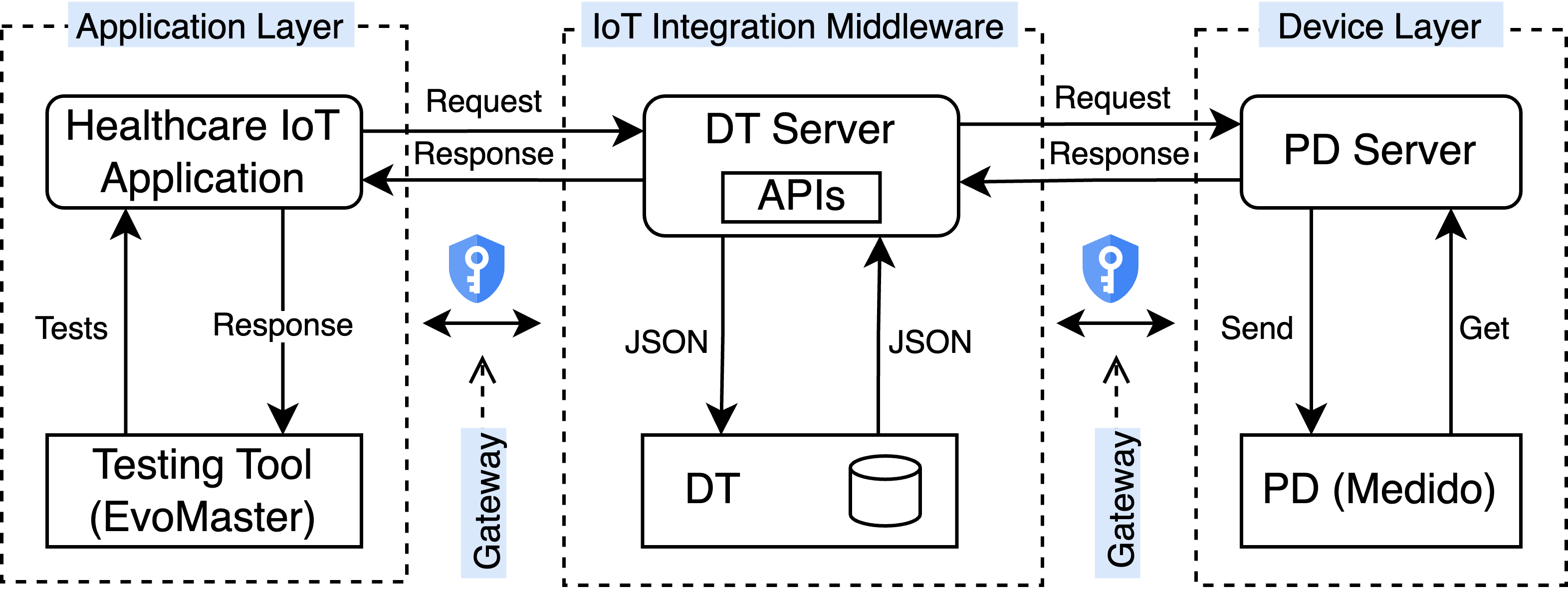}
\caption{Evaluation setup with a healthcare IoT application, testing tool, DTGen component, and physical device (PD).} 
\label{fig:setup}
\end{figure}

\subsection{Experiment Setup and Execution}
\subsubsection{Real-world Case Study.}
We used Medido~\cite{medido} medicine dispenser integrated with a healthcare IoT application provided by Oslo City as a part of the experiment apparatus.  
Medido is a multi-featured automatic medicine dispenser that provides various functionalities to stakeholders. It enables healthcare specialists and caretakers to personalize device settings, including language, alarm, and medication plans. Its key operations are loading medication plans from the healthcare IoT application, following a plan to dispense medicine at the specified time, and notifying concerned healthcare specialists/caretakers regarding missed doses and medicine dispense problems.

\subsubsection{Setup.}
Figure~\ref{fig:setup} presents an overview of the evaluation setup. 
At \textit{application layer}, we used a healthcare IoT application provided by Oslo City as a part of the experiment apparatus. 
We used a REST API testing tool, namely EvoMaster~\cite{arcuri2019restful}. 
At \textit{device layer}, we utilized Medido~\cite{medido} medicine dispenser supplied by Oslo City. 
Medido is connected to its server, which manages communication with the device. 
For the \textit{IoT integration middleware}, we generated DTs of Medido using model-based and ML approaches, namely MBDT and MLDT, respectively. 
For MBDT, we modeled the domain, constraints, and state machines of Medido. 
For MLDT, we performed a pilot experiment to determine the NN architecture and hyperparameters. 
The resulting NN has two hidden layers with dimensions eight and four, {\fontfamily{qcr}\selectfont dropout rate=0.2}, and a Sigmoid activation function. 
The remaining hyperparameters are {\fontfamily{qcr}\selectfont learning rate=0.01}, 
{\fontfamily{qcr}\selectfont epochs=3000}, {\fontfamily{qcr}\selectfont optimizer=Adam}, and {\fontfamily{qcr}\selectfont loss function=Cross-entropy}.
For both MBDT and MLDT, we created DT APIs and configured a DT Server. 
We used JSON as a data interchange format for communication with DTs. 

For each RQ, we executed EvoMaster to generate test data and sent this data to both MBDT and MLDT, as well as to Medido. 
EvoMaster was configured to run for a duration of two hours at a rate of 100 API requests per minute. 
In each execution, we collected responses from MBDT, MLDT, and Medido. 
Using responses, we analyzed the fidelity (for RQ1) of MBDT and MLDT with Medido based on the similarity in responses.  
For RQ2, we generated 100 MBDTs and 100 MLDTs and run them in different batch sizes, i.e., 10, 20, 30, ..., 100. 
We sent test data to all MBDTs and MLDTs running in different batches and to Medido and compared the responses of all DTs.  
For RQ3, we executed each machine-dependent automated step 10 times on one machine to analyze the average time.

\subsubsection{Execution.}
We ran experiments using a machine with a macOS operating system, an 8-core CPU, and 24 GB RAM.

\subsection{Metrics and Statistical Tests}
We analyze DTs' fidelity in terms of their operating similarity with PD. 
For this purpose, we used the Cosine similarity measure. 
To statistically analyze DTs' fidelity, we also used the Wilcoxon test signed rank and Cliff's Delta ($\delta$) with a significance level ($\alpha$) 0.05, following guidelines by Arcuri and Briand~\cite{arcuri2011practical}.

\subsection{Results and Discussion}
Following, we discuss results corresponding to each RQ. 

\subsubsection{RQ1: DT Fidelity.}
Table~\ref{tab:rq1} shows results of MBDT and MLDT fidelity in terms of their similarity with Medido PD. 
The similarity of MBDT and MLDT with Medido PD is $\approx$94\% and $\approx$95\%, respectively, indicating both types of DTs have close operational resemblance with PD. 
The Wilcoxon test p-values are greater than $\alpha$ for both MBDT and MLDT. 
This shows there is no significant difference between MBDT and PD and MLDT and PD. 
The effect size analysis indicates that the difference magnitude is negligible for both MBDT and MLDT. 
This indicates a high operational similarity of MBDT and MLDT with their corresponding Medido PD.

\begin{rqres}
  DTs generated using model-based and ML approaches have fidelity compared to Medido. 
\end{rqres}

\begin{table}[!htb]
  \floatsetup{floatrowsep=qquad, captionskip=4pt}
  \ttabbox%
    {\begin{tabularx}{\textwidth}{l *{2}{>{\centering\arraybackslash}X}}
      \toprule
      & \textbf{MBDT} & \textbf{MLDT} \\
      \cmidrule(lr){2-2}\cmidrule(lr){3-3}
      \textbf{Cosine Similarity} & 94.05\% & 95.54\% \\
      \addlinespace
      \textbf{Wilcoxon Test ($p-value$)}& 1.0 & 1.0\\
      \addlinespace
      \textbf{Effect Size ($\delta$)} & -0.12 (negligible) & -0.09 (negligible)\\
      \bottomrule
      \end{tabularx}}
    {\caption{RQ1: Fidelity of MBDT and MLDT compared with Medido PD}
      \label{tab:rq1}}
\end{table}%

\subsubsection{RQ2: DTGen Component Scalability.}
\begin{figure}
\centering
\includegraphics[width=\linewidth]{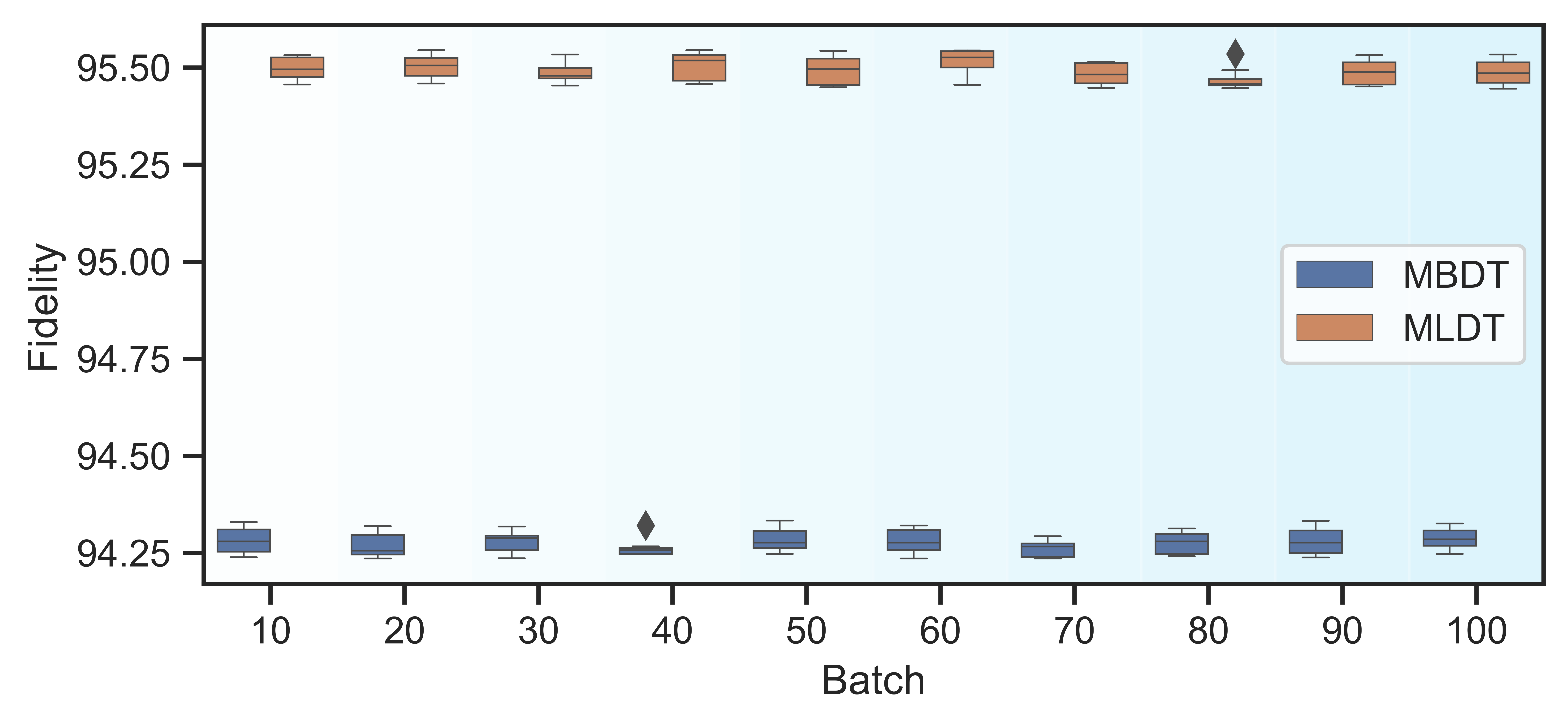}
\caption{RQ2: Scalability of multiple MBDTs and MLDTs running in different batch sizes} 
\label{fig:rq2}
\end{figure}

Figure~\ref{fig:rq2} shows boxplots for fidelity comparison of MBDTs and MLDTs running in different batch sizes, i.e., 10, 20, 30, ..., 100, highlighted with increasing gradient color. 
It can be observed that the median fidelity level of MBDTs is approximately 94\% across different batches. 
Similarly, MLDTs' median fidelity is approximately 95.5\% for varying batch sizes. 
It is worth noticing that the fidelity of different batches of MBDTs and MLDTs with their corresponding Medido PD is consistent with the increase in DTs number. 
Furthermore, the fidelity values reported in Table~\ref{tab:rq1} for one MBDT and MLDT are in order with the median fidelity values observed while operating multiple MBDTs and MLDTs in various batch sizes. 
This indicates that operating different numbers of DTs has nearly identical fidelity as the number of DTs grows.

\begin{rqres}
  The fidelity of 100 DTs (MBDTs and MLDTs) is consistent in all batch sizes, highlighting that the DTGen component is scalable. 
\end{rqres}

\subsubsection{RQ3: Time Cost.}
Table~\ref{tab:mbcost} and~\ref{tab:mlcost} present the estimated time cost involved in various steps of creating MBDT and MLDT, respectively. 
To generate MBDT, modeling domain, constraints, and behavior requires significant time, approximately 1-2 hours, which can vary depending on the modelers' experience. 
Providing input configurations may take 2-5 minutes. 
It is important to note that generating one or more MBDTs is an automated step and only takes time in milliseconds. 
The time required to generate one MBDT and 100 MBDTs is, on average, $\approx$108 and $\approx$229 milliseconds, respectively. 

For MLDT, the data collection step requires integrating a testing tool and compiling data during execution. 
This semi-automatic step takes 2-3 hours to collect sufficient data set for training. 
In case more data set is required, this can take more time. 
Configuring ML involves identifying suitable NN architecture and hyperparameters search that can take approximately 1-2 hours. 
Training MLDT with 3000 epochs takes 3 minutes and 15 seconds, and with 5000 epochs it takes 5 minutes and 29 seconds. 
In our experiments, 3000 epochs are sufficient to obtain a well-trained model; however, training with higher epochs is not high. 
The time required to generate one MLDT and 100 MLDTs is, on average, $\approx$4 and $\approx$38 milliseconds, respectively. 
The automatic steps' time cost varies with machine specifications, thus requiring further empirical study involving machines with different specifications. 

\begin{rqres}
  DTs generated using ML require fewer manual steps and more time compared to model-based DT. ML approach with higher time cost generates DTs with higher fidelity compared to model-based DTs. 
\end{rqres}

\begin{table}[!htb]
  \floatsetup{floatrowsep=qquad, captionskip=4pt}
  \ttabbox%
    {\begin{tabularx}{\textwidth}{l *{3}{>{\centering\arraybackslash}X}}
      \toprule
      & \textbf{Steps} & \textbf{Automation} & \textbf{Time} \\
      \cmidrule(lr){2-2}\cmidrule(lr){3-3}\cmidrule(lr){4-4}
      \textbf{Modeling}&Domain model, OCL constraints, state machines& Manual & 1-2 hrs \\
      \addlinespace
      \textbf{Inputs Configs.}&Device properties, server settings, API mapping& Manual & 2-5 min\\
      \addlinespace
      \textbf{DT Generation}& Create executable models and DT storage, operate DT & Automatic & 108.428 ms (avg)\\
      \addlinespace
      \textbf{100 DTs }& Create 100 DTs & Automatic & 228.931 ms (avg)\\
      \bottomrule
      \end{tabularx}}
    {\caption{RQ3: MBDT time cost considering main/sub-steps, automation level, and average time}
      \label{tab:mbcost}}
\end{table}%

\begin{table}[!htb]
  \floatsetup{floatrowsep=qquad, captionskip=4pt}
  \ttabbox%
    {\begin{tabularx}{\textwidth}{l *{3}{>{\centering\arraybackslash}X}}
      \toprule
      & \textbf{Steps} & \textbf{Automation} & \textbf{Time} \\
      \cmidrule(lr){2-2}\cmidrule(lr){3-3}\cmidrule(lr){4-4}
      \textbf{Data Collection}&Setup and run testing tool, compile data from responses & Semi-Automatic & 2-3 hrs \\
      \addlinespace
      \textbf{ML Configs.}& NN design, tune hyperparameters& Semi-automatic & $\approx$1-2 hr\\
      \addlinespace
      \textbf{Training}&Train and save NN model& Automatic & 3m 15s - 5m 29s\\
      \addlinespace
      \textbf{DT Generation} &Load NN model, create storage, operate DT& Automatic & 4.551 ms (avg)\\
      \addlinespace
      \textbf{100 DTs}&Create 100 DTs& Automatic & 38.208 ms (avg)\\
      \bottomrule
      \end{tabularx}}
    {\caption{RQ3: MLDT time cost considering main/sub-steps, automation level, and average time}
      \label{tab:mlcost}}
\end{table}%

\subsection{Threats to Validity}
To minimize potential \textbf{external validity} threats, we evaluated \mbox{HITA}'s DTGen component utilizing a Medido medicine dispenser integrated with a real-world healthcare IoT application. Medido is a widely used and a good representative medicine dispenser. We intend to add more medical devices to conduct a large-scale evaluation in the future. 
To reduce \textbf{internal validity} threats, we carefully designed experiments based on device and API documentation from Oslo City. We conducted sessions with industry practitioners to showcase the setup and get their feedback. Apart from initial configurations, the MBDT generation approach involves no further parameter tuning during execution. 
Regarding the MLDT experiment, we conducted pilot experiments for hyperparameter search, ensuring careful consideration of hyperparameters. 
To handle \textbf{construct and conclusion validity} threats, we used the Cosine similarity measure, the Wilcoxon test signed rank, and Cliff's Delta. We used the suggested significance level~\cite{arcuri2011practical} for statistical analysis. 
In the time cost analysis (RQ3), we executed each automated step 10 times for both MBDT and MLDT to calculate the average time, ensuring robustness in our analysis.

\section{Experiences and Lessons Learned}\label{learnings}
Following we outline our experience and lessons learned while developing \mbox{HITA} work products and analyzing them through experiments. 

\subsection{DTs Role in Test Infrastructure}
System-level testing of healthcare IoT applications requires different medical devices in the loop.  
Each type of medical device from a different vendor is linked to a web server that has certain constraints on maximum allowed requests. 
The test generation and execution process involves sending several requests to medical devices through a healthcare IoT application. 
This leads to the blocking of service or the damaging of a medical device. 
Further, testing with hundreds of such devices is costly and not a practical option. 
Based on such experiences from Oslo City, we propose the idea of using DTs in place of physical medical devices to enable testing with multiple digital representations of physical devices. 
Thus, DTs have an important role in this regard. 
DTs with dedicated \textit{DT Server} and APIs eliminate the risk of service blockage or device damage. 
Virtually representing physical devices, DTs are a scalable and cost-effective solution. 
Our experiments with 100 DTs in different batches (i.e., 10, 20, 30, ..., 100) indicated scalability, heterogeneity, and cost-effectiveness of the DTGen component.  

\subsection{Trade-off between model-based and ML DTs}
The model-based approach for the automated generation of DTs requires creating domain models of medical devices capturing devices' structural aspects and modeling behavioral aspects of medical devices using executable state machines. 
Several modeling tools (e.g., IBM RSA and Papyrus) are available for this purpose. 
Test engineers need to have a fundamental level of familiarity with any of the modeling tools.
Models developed in this way involve a one-time effort and can be reused for testing multiple evolution phases of SUT~\cite{sartaj2021testing,sartaj2020cdst}. 
In the case of adding or upgrading medical devices, only the domain model and executable state machines need to be fine-tuned. 

For generating DTs using the ML approach, training data needs to be generated with medical devices in the loop. 
A key consideration for generating training data is a device's request processing capability. 
For example, if a device takes two seconds to process a request, sending many requests without delay may damage the device. 
With training data collected and preprocessed, identifying suitable neural network architecture and hyperparameters for training requires several experiment trials. 
These steps are largely automated; nevertheless, for each new/upgraded device, these steps must be repeated. 

Using model-based or ML approaches for DT generation depends on the industrial application context. 
The model-based approach requires more manual steps compared to the ML approach. 
However, device data is a fundamental requirement for using the ML approach which needs to be generated using test devices. 
Since model-based and ML DTs have nearly similar fidelity, either approach can be employed with test infrastructure. 

\subsection{Fidelity Evaluation of DTs}
While utilizing DTs of physical devices, an important consideration is the fidelity of DTs corresponding to physical twins. 
For this purpose, we empirically evaluated the fidelity of model-based and ML DTs (up to 100) in terms of their functional similarities with a physical medicine dispenser (Medido). 
The results highlighted the functionality of DTs was almost similar to medicine dispensers. 
Moreover, fidelity evaluation in terms of internal behaviors is challenging due to limited access to internal operations of physical medicine dispensers.

\subsection{Testing with Third-party Applications}
We experimented with testing the REST API of a healthcare IoT application (SUT) connected to different third-party applications~\cite{sartaj2023testing}. 
In our experiments, we observed that API failures of third-party applications during test execution pose a challenge in pinpointing faults within the SUT. 
We also noticed that services provided by third-party applications often become unavailable after numerous test executions, causing a hindrance to the rigorous testing of healthcare IoT applications. 
Hence, the creation of test stubs for third-party applications, as suggested in \mbox{HITA}, appears to be a viable solution.

\subsection{Domain-specific Testing Strategies}
Our experiments with REST API testing highlighted the need for domain-specific testing strategies for healthcare IoT applications~\cite{sartaj2023testing}. 
We analyzed that automated realistic test data generation is a challenging and open research problem. 
For example, automatically generating a valid medication plan for a patient is not a simple task. 
Generating a valid medication plan requires information regarding the start date, dose intake, number of days to take medicines, number of doses, and the total number of medicines allowed in a roll of a medicine dispenser. 
This involves understanding domain properties related to medications and the context of a medicine dispenser.
There is still a need for domain-specific testing strategies.

\subsection{Intelligent Test Generation Technique}
Healthcare IoT applications commonly have a two-way communication mechanism with different medical devices and third-party applications. 
Several scenarios require an integrated medical device or third-party application to initiate the first step of the process. 
Automatically generating test cases for such scenarios is challenging. 
For instance, the steps to assign an alert (received from a patient) to concerned personnel include: (i) the patient's medical device generates an alert, (ii) the alert is received as an unassigned alert, (iii) identify an appropriate person (doctor, nurse, caretaker, etc.) to assign the alert, and (iv) assign the alert with notification to health authorities. 
An alert should be generated beforehand to test the alert-assigning scenario. 
This requires an intelligent technique for automated test case generation since \mbox{HITA} is designed for creating test infrastructure. 

\subsection{Test Optimization}
Testing an industrial healthcare IoT application in production and a rapid-release environment requires a designated time budget for test generation and execution. 
Executing a maximum number of test cases with the aim of rigorous testing for each release is desirable but not feasible, even using test stubs and digital twins. 
An approach for generating and executing optimized test cases is necessary to ensure the dependability of healthcare IoT applications within a given time frame.

\section{Conclusion and Future Work}\label{conclusion}
In this paper, we presented real-world architectural work in collaboration with Oslo City's healthcare department. 
We introduced \mbox{HITA} -- a test infrastructure architecture to facilitate automated system-level testing of healthcare IoT applications, with design considerations aligned to Oslo City's healthcare department requirements. 
We evaluated \mbox{HITA}'s DTGen component by creating DTs of a Medido medicine dispenser using model-based and ML approaches. 
Our evaluation focused on analyzing the fidelity of DT created using both approaches, assessing the scalability of the DTGen component when operating 100 DTs in different batch sizes, and the time cost involved in creating DTs. 
Results show that the fidelity of DTs created using model-based and ML approaches is 94\% and 95\%, respectively. 
Results with 100 DTs also show that the DTGen component is scalable. 
Moreover, results indicate that DTs generated with the ML approach have fewer manual steps and higher time costs compared to model-based DTs. 
Finally, we presented experience and lessons learned based on experiments conducted with work products of \mbox{HITA} that are valuable for industry practitioners working in a similar domain. 
Our architecture, findings, and lessons learned are generalizable to various IoT-based systems such as activity/fitness trackers, smart homes, and smart security systems.

In the future, we plan to extend the DTGen component and create DTs of different types of medical devices. We also intend to implement \mbox{HITA}'s TSGen component supporting third-party applications. Next, we plan to develop domain-specific testing strategies focusing on GUI testing, REST API testing, and test optimization.

\subsubsection*{Acknowledgements.}
This work is a part of the WTT4Oslo project (No. 309175) funded by the Research Council of Norway.
All the experiments reported in this paper are conducted in a laboratory setting of Simula Research Laboratory; therefore, they do not by any means reflect the quality of services Oslo City provides to its citizens. 
Finally, we would like to acknowledge Kjetil Moberg for providing feedback on the initial version of this paper. 

%
%
%
\bibliographystyle{splncs04}
\bibliography{refs}
\end{document}